\documentclass[twocolumn]{jpsj3}
\usepackage{graphicx,amsmath,amssymb,bm}
\usepackage{braket}

\allowdisplaybreaks

\usepackage{color}
\definecolor{Green}{rgb}{0,0.7,0}
\newcommand{\e}{ {\rm e}}
\newcommand{\bk}{\bm{k}}

\newcommand{\bkD}{\bk_{\rm D}}

\newcommand{\ET}{$\alpha$-ET$_2$I$_3$ }

\newcommand{\tx}{\bk \bm{a}}
\newcommand{\ty}{\bk \bm{b}}
\newcommand{\tz}{\bk \bm{c}}

\newcommand{\kx}{k_x}
\newcommand{\ky}{k_y}
\newcommand{\kz}{k_z}
\newcommand{\HH}{{\rm H}}
\newcommand{\LL}{{\rm L}}

\newcommand{\bX}{\bar{X}}
\newcommand{\bY}{\bar{Y}}
\newcommand{\bZ}{\bar{Z}}

\newcommand{\m}{+}
\newcommand{\p}{-}
\newcommand{\hg}{{\bf G}/2}
\newcommand{\ho}{{\rm H}}
\newcommand{\lo}{{\rm L}}

\begin{document}


\title{
Electronic Structure
of a Single-Component Molecular Conductor [Pd(dddt)$_2$]
(dddt = 5,6-dihydro-1,4-dithiin-2,3-dithiolate)
under High Pressure
}
\author{
Reizo Kato,$^{1}$\thanks{E-mail: reizo@riken.jp}
 Hengbo Cui,$^{1}$
  Takaaki Minamidate,$^{1}$
Hamish H.-M. Yeung,$^{2}$\thanks{Present address: School of Chemistry, University of Birmingham, Edgbaston, Birmingham, B15 2TT, UK}
 and
Yoshikazu Suzumura$^{3}$\thanks{E-mail: suzumura@s.phys.nagoya-u.ac.jp}
}
\inst{
$^1$
RIKEN, 2-1 Hirosawa, Wako-shi, Saitama 351-0198, Japan \\
$^2$
Inorganic Chemistry Laboratory, University of Oxford,
South Parks Road, Oxford OX1 3QR, UK \\
$^3$
Department of Physics, Nagoya University, Chikusa-ku, Nagoya 464-8602, Japan \\%
}

\recdate{August 24, 2020; accepted October 2, 2020}
\abst{
We examined the high-pressure electronic structure of a single-component molecular conductor [Pd(dddt)$_2$] (dddt = 5,6-dihydro-1,4-dithiin-2,3-dithiolate) at room temperature,
 on the basis of the crystal structure determined by single-crystal synchrotron X-ray diffraction measurements at 5.9 GPa.
The monoclinic unit cell contains four molecules that form two crystallographically independent molecular layers.
A tight-binding model of an 8 $\times$ 8 matrix Hamiltonian gives an electronic structure as a Dirac electron system.
The Dirac point describes a loop within the first Brillouin zone, and a nodal line semimetal is obtained.
The noticeable property of the Dirac cone with a linear dispersion is shown
by calculating the density of states (DOS).
The Dirac cone in this system is associated with the crossing of highest occupied molecular orbital (HOMO) and lowest unoccupied molecular orbital (LUMO) bands, which originates from the direct interaction between
different molecular layers.
This is a newly found mechanism in addition to
the indirect interaction [J. Phys. Soc. Jpn., {\bf 86}, 064705 (2017)].
 The Dirac points emerge as a line when the HOMO and LUMO bands meet on the surface and the HOMO--LUMO couplings are absent.
Such a mechanism is verified using a reduced model of a 4 $\times$ 4 matrix Hamiltonian.
The deviation of the band energy ($\delta E$)
at the Dirac point from the Fermi level is very small
($\delta E < $ 0.4meV).
 The nodal line is examined by calculating
 the parity of the occupied band eigenstates at time reversal invariant momentum (TRIM), which  shows that the topological number is 1.
}


\maketitle

\section{Introduction}
Molecular conductors have simple and clear electronic structures where a simple extended H\"uckel tight-binding (TB) band picture is applicable.
\cite{r1}
This is mainly because only one kind of frontier molecular orbital (HOMO or LUMO) in each molecule contributes to the formation of a conduction band in conventional molecular conductors,
where HOMO and LUMO denote
highest occupied molecular orbital and
lowest unoccupied molecular orbital, respectively.
In recent years, however, the number of molecular conductors that cannot be categorized as such a single-orbital system has been increasing.
This means that we should expand our perception toward a multi-orbital system where more than two molecular orbitals in the same molecule contribute to electronic properties and the orbital degree of freedom plays an essential role.

A typical example of the multi-orbital system is a single-component molecular conductor.
Molecules usually have a closed-shell electronic structure, which is the reason why they are stable in an isolated state.
Therefore, it was believed that neutral closed-shell molecules do not self-assemble to form a metallic bond in a crystal and, therefore, the electron transfer between the frontier molecular orbital and other chemical species is indispensable for a metallic molecular crystal.
Resultant molecular metals are no longer single-component, but they include other cations or anions to maintain the charge neutrality. If the energy difference between HOMO and LUMO is sufficiently small, however, the fully occupied HOMO band and the empty LUMO band can overlap, and an intramolecular electron transfer leads these bands to partially filled states.
This idea has been confirmed by the observation of electron and hole Fermi surfaces in an ambient-pressure single-component molecular metal [Ni(tmdt)$_2$]
(tmdt = trimethylenetetrathiafulvalenedithiolate) by detecting the de Haas--van Alphen effect.\cite{r2,r3}
 After this breakthrough, various single-component molecular conductors have been developed using metal dithiolene complexes.\cite{r4}
In metal dithiolene complexes with a planar central core, the HOMO is destabilized owing to the absence of the contribution from the metal d orbitals and the HOMO--LUMO gap is small ($<1$ eV) in general.\cite{r5}
Even in the case of metal dithiolene complexes, however, the HOMO and LUMO bands are mostly separated from each other, and a metallic state rarely emerges at ambient pressure. In a molecular crystal with a soft lattice, the application of high pressure can effectively enhance intermolecular transfer integrals and thus induce the overlap of the HOMO and LUMO bands. Indeed, an increasing number of single-component molecular metals have been found under high pressure.\cite{r6,r7}%
Notably, an improvement of the diamond anvil cell (DAC) technique that provides high-quality quasi hydrostatic pressure drove the research forward.\cite{r8}
Superconductivity in a single-component molecular crystal was also achieved under high pressure generated by DAC.\cite{r9}
 Thus, we realized a metallic/superconducting state in single-component molecular crystals.
In these single-component molecular systems that we now focus on, molecules maintain their original molecular properties even in the high-pressure metallic state, and it is anticipated that the emergence of a metallic state itself can be well understood in the framework of the conventional TB band theory.
It should be recognized that the deeper goal of the physical research on single-component molecular conductors is not just to obtain a conventional metallic state.
What is really being put to the test is the possibility of unique physical properties that are built in the multi-orbital system.

  In this sense, the discovery of a nodal line semimetal state in a single-component molecular conductor [Pd(dddt)$_2$]
(dddt = 5,6-dihydro-1,4-dithiin-2,3-dithiolate) under high pressure has paved the way for the development of multi-orbital molecular conductors.
\cite{r10,r11,r12,r13,Liu2018,r14}
Indeed, after this discovery, a semimetal with open nodal lines
 has been found in another single-component molecular conductor [Pt(dmdt)$_2$] (dmdt = dimethyltetrathiafulvalenedithiolate).\cite{Zhou2019,Kato2020_JPSJ, Kobayashi2020}
Nodal line semimetals where the conduction and valence bands touch each other
along a line in the three-dimensional Brillouin zone have aroused broad interest owing to the possibility of topologically nontrivial states.
\cite{
r15,r16,Kim_Rappe_NLS_2015,Yamakage2016_JPSJ,Fang2016,Suzumura_Yamakage_JPSJ,Yang2018,Hirayama2017,Hirayama2018,Bernevig2018
}
The crystal of the metal dithiolene complex [Pd(dddt)$_2$] is an insulator at ambient pressure. The application of hydrostatic pressure using the DAC technique suppressed resistivity and activation energy. The temperature-independent resistivity observed at 12.6 GPa triggered theoretical studies using first-principles calculations based on the density functional theory (DFT).
The energy band structure for the optimized high-pressure structure indicates the emergence of the Dirac cones at 8 GPa,
which is consistent with the zero-gap behavior observed in the resistivity measurement.
The TB model based on extended H\"uckel molecular orbital calculations revealed that the Dirac cone formation is associated with the multi-orbital character, and the Dirac point describes a loop in the three-dimensional Brillouin zone.
The small deviation of the energy on the loop from the Fermi level gives hole and electron pockets, which means that the system is a nodal line semimetal.
Although an unexpected relationship between the single-component molecular conductor and the nodal line semimetal has been disclosed, a concern is the absence of X-ray structural data of the [Pd(dddt)$_2$] crystal determined under high pressure.
The energy band calculations were based only on theoretically optimized cell parameters and atomic coordinates.

Recently, crystal structures of [Pd(dddt)$_2$] at several pressures have been determined by single-crystal synchrotron X-ray diffraction measurements, details of which will be reported elsewhere. In this article, we examine the nodal line semimetal state of [Pd(dddt)$_2$] at 5.9 GPa at room temperature on the basis of the determined crystal structure.
This paper is organized as follows.
In Sect. 2, we describe the TB model used in this work.
In Sect. 3, first, the band structure is shown with a detailed description of
  the mechanism of the Dirac cone formation and the resultant nodal line. Next, we present the density of states (DOS) and parity at TRIM, which are relevant to the nodal line. The conclusion is given in Sect. 4.

  \section{TB model}
  We carried out synchrotron X-ray diffraction measurements at several pressures and found that there is no marked structural phase transition up to 10.6 GPa. The cell volume measured at 5.9 GPa (1152.8 \AA$^3$ ) is close to that obtained by the DFT calculation for the 8 GPa structure (1147.5 \AA$^3$), which is the reason why we focus on the 5.9 GPa structure. The determined cell parameters and atomic coordinates are listed in Appendix A.
  Figure \ref{fig:fig1} shows the molecular arrangement and intermolecular couplings in the [Pd(dddt)$_2$] crystal.
  The unit cell contains four molecules (1, 2, 3, and 4), and each central Pd atom is located at the inversion center.
  The [Pd(dddt)$_2$] molecules uniformly stack along the $b$-axis that is perpendicular to the $ac$ plane.
  Crystallographically equivalent molecules form two types of layers, layer 1 (molecules 1 and 3)
  and layer 2 (molecules 2 and 4), both of which are parallel to the $ab$ plane.

\begin{figure*}[tb]
\begin{center}
  \includegraphics[width=0.9 \linewidth]{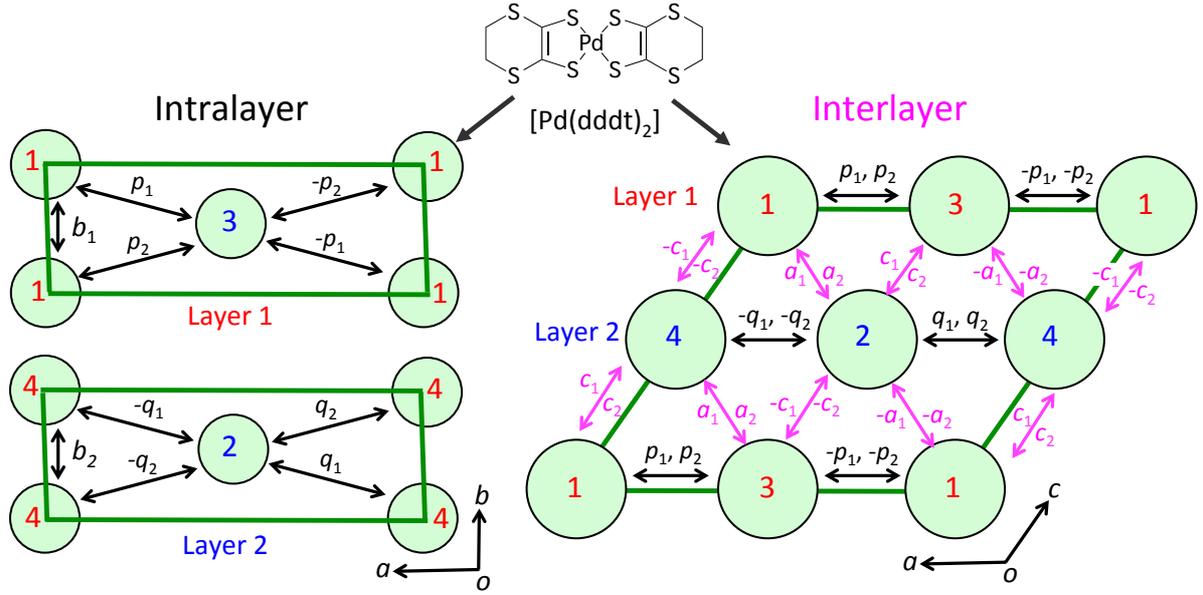}
\end{center}
\caption{
(Color online)
  Molecular arrangement and intermolecular couplings
 that explain crystal structure of [Pd(dddt)$_2$].\cite{r11}
}
\setlength\abovecaptionskip{0pt}
\label{fig:fig1}
\end{figure*}

\begin{table}
\caption{
HOMO--HOMO (H--H), LUMO--LUMO (L--L), and HOMO--LUMO (H--L) transfer energies (meV) at 5.9 GPa.

}
\begin{center}
\begin{tabular} {ccccc}
\hline\noalign{\smallskip}
 $ $          & H--H     & L--L   & H--L    & \\
\hline\noalign{\smallskip}
 $ b1 $       & $209.3$   & $-1.9$   & $-51.2$ & (stacking)  \\
 $ p1(p) $    & $28.1$    &  $-12.4$  & $19.9$  & Layer 1   \\
 $ p2 $       & ---     &  ---  & $17.1 $  &      \\
\noalign{\smallskip}\hline\noalign{\smallskip}
 $ b2 $       & $49.9$   &  $-80.4$   & $-67.2$  &  (stacking)  \\
 $ q1(q) $    & $10.8$    &  $8.1$  & $9.3$  & Layer 2   \\
 $ q2 $       & ---     &  ---  & $9.2$  &      \\
\noalign{\smallskip}\hline\noalign{\smallskip}
 $ a1 $       & $-28.2$   &  $14.6$   & $-20.1$  &    \\
 $ a2 $       & $2.2$    &  $1.3$  & $-1.7$  &  Interlayer   \\
 $ c1 $       & $15.4$     &  $12.7$  & $14.1$  &      \\
 $ c2 $       & $-3.9$     &  $15.8$  & $-11.8$  &

\\
\noalign{\smallskip}\hline
\end{tabular}
\end{center}
\label{table1}
\end{table}

We construct a TB model for [Pd(dddt)$_2$] using frontier molecular orbitals of four molecules in the unit cell, HOMOs (H1, H2, H3, H4) and LUMOs (L1, L2, L3, L4).

The TB model Hamiltonian is given by
\begin{eqnarray}
H_{\rm TB}  &=& \sum_{i,j=1}^{N} \sum_{\alpha,\beta} t_{i,j;\alpha, \beta} \ket{i, \alpha}\bra{j, \beta} \nonumber \\
    & =&
   \sum_{\bk} \sum_{\alpha,\beta} h_{\alpha, \beta}(\bk)\ket{\bk, \alpha}\bra{\bk, \beta} \nonumber \\
  & =&
    \sum_{\bm{k}} \ket{\Phi(\bm{k})} \hat{H}(\bm{k}) \bra{\Phi(\bm{k})}\; ,
 \label{eq:eq1}
 \end{eqnarray}
where
 $t_{i,j;\alpha, \beta}$ are transfer energies between nearest-neighbor sites and $\bra{i, \alpha}$ is a state vector.
$\alpha, \beta$ = H1, H2, $\cdots$, L3, and L4.
  $h_{\alpha, \beta}(\bk)$ denotes a Fourier transform of
 $t_{i,j;\alpha, \beta}$ with  a complex conjugate relation  $h_{\alpha, \beta}(\bk) = \overline{h_{\beta, \alpha}(\bk)}$, where
 $\bk= \kx a^* + \ky b^* +\kz c^* \equiv (\kx, \ky, \kz) $, and $2\pi\kx=\tx,
2\pi\ky=\ty$, and $2\pi\kz=\tz$.
 $\hat{H}(\bm{k})$ is an 8 $\times$ 8 matrix Hamiltonian, where
$h_{\alpha,\beta}  =  \left( \hat{H}(\bk) \right)_{\alpha,\beta}$
and $\bra{\Phi(\bm{k})} = (\bra{\HH1},\bra{\HH2},\bra{\HH3},\bra{\HH4}, \bra{\LL1}, \bra{\LL2}, \bra{\LL3}, \bra{\LL4}).$
In terms of
$X = \e^{i\tx}$,
$Y = \e^{i\ty}$,  and
$Z = \e^{i\tz}$,
matrix elements $h_{\alpha,\beta}(\bk)$ are given in Appendix B.
The energy difference between HOMO and LUMO is taken as $\Delta E$ =
0.696 eV  to reproduce the energy band obtained by the DFT calculation.
Interlayer and intralayer transfer energies in matrix elements $h_{\alpha,\beta}(\bk)$ are shown in Table \ref{table1}, which are estimated by the extended H\"uckel method.
The interlayer transfer energies in the $c$ direction are given
by $a$ (molecules 1 and 2, and molecules 3 and 4)
and $c$ (molecules 1 and 4, and molecules 2 and 3).
The intralayer transfer energies parallel to the $ab$ plane are given by
$p$ (molecules 1 and 3 ), $q$ (molecules 2 and 4), and $b$.
These transfer energies are classified into HOMO--HOMO (H), LUMO--LUMO (L), and HOMO--LUMO (HL).

These transfer energies are  rather different from those of the previous model based on the theoretically optimized structure at 8 GPa.
Notably, there are significant contributions from
the direct interlayer HOMO--LUMO couplings, which provide the elements
$ h_{\rm H1, L2}, h_{\rm H1, L4}, h_{\rm H3, L2}$, and $h_{\rm H3, L4}$ and those being
the complex conjugate elements.
In the previous model in Ref. \citen{r11}, however,
these elements are negligibly small
 and thus discarded. The indirect interlayer HOMO--LUMO couplings play an important role in the Dirac cone formation instead.
 This is obtained by a second-order perturbation in terms of
 the intralayer HOMO--LUMO and interlayer LUMO--LUMO or HOMO--HOMO
couplings, e.g., a combination of
 $h_{\rm H1, L3}$ and $h_{\rm L3, L2}$.
The direct interlayer HOMO--LUMO couplings also give the off-diagonal elements
 of a reduced model of a 4 $\times$ 4 Hamiltonian, which we will discuss later.
Since the symmetry of the HOMO (H) (LUMO (L))
 is odd (even) with respect to the
 Pd atom, the matrix element of H--L (H--H and L--L)
 is the odd (even) function
 with respect to $\bm{k}$.

The energy band $E_j(\bk)$
 and the wave function $\Psi_j(\bk)$, $(j = 1, 2, \cdots, 8)$
 are calculated from
\begin{equation}
\hat{H}(\bm{k}) \Psi_j(\bk)
 = E_j(\bk) \Psi_j(\bk) \; ,
\label{eq:eq2}
\end{equation}
 where $E_1 > E_2 > \cdots > E_8$.
Since the energy band formed by fully occupied HOMOs and empty LUMOs is half-filled,
 we examine the gap defined by
\begin{equation}
E_g(\bk) = {\rm min} (E_4(\bk)-E_5(\bk)) \; ,
\label{eq:eq3}
\end{equation}
for all $\bk$ in the Brillouin zone.
The Dirac point $\bk_D$ is obtained from $E_g (\bkD)= 0$, which  leads to a nodal line.

 We also examine the nodal line using an effective Hamiltonian given by
\begin{eqnarray}
 {H_{\rm eff}(\bm{k})}
 &=&
\begin{pmatrix}
f_0(\bk)+f_3(\bk) &  f_2(\bk)  \\
f_2(\bk) & f_0(\bk)- f_3(\bk)
\end{pmatrix}  \; , \nonumber \\
\label{eq:eq4}
\end{eqnarray}
where $f_0(\bk)$, $f_3(\bk)$, and $f_2(\bk)$ are calculated as follows.
The Hamiltonian is divided into three 8 $\times$ 8 matrices as
\begin{eqnarray}
\hat{H}= \hat{H}_{\ho-\ho} + \hat{H}_{\lo-\lo}+ \hat{H}_{\ho-\lo} \; ,
\label{eq:eq5}
\end{eqnarray}
where
 $[\hat{H}_{\ho-\ho}]_{\alpha,\beta} = h_{\alpha,\beta}$
    with  $\alpha, \beta = {\rm H1}, \cdots {\rm H4}$ and 0 otherwise,
 $[\hat{H}_{\lo-\lo}]_{\alpha,\beta} = h_{\alpha,\beta}$
    with   $\alpha, \beta = {\rm L1}, \cdots {\rm L4}$ and 0 otherwise, and
 $[\hat{H}_{\ho-\lo}]_{\alpha,\beta} = h_{\alpha,\beta}$
  with  $ \alpha=  {\rm H1} \cdots,{\rm  H4}$
       (or ${\rm L1}, \cdots {\rm L4}$) and
  $ \beta ={\rm L1}, \cdots {\rm L4}$
        (or ${\rm H1} \cdots,{\rm  H4}$) and 0 otherwise.
We define $E_{\rm H}(\bk)$ ($E_{\rm L}(\bk)$) as the maximum (minimum)
eigenvalue of HOMO (LUMO), which is obtained from
 $ \hat{H}_{\ho-\ho}\ket{\Psi_{\rm H}} = E_{\rm H}(\bk)\ket{\Psi_{\rm H}}$ and
 $ \hat{H}_{\lo-\lo}\ket{\Psi_{\rm L}} = E_{\rm L}(\bk) \ket{\Psi_{\rm L}}$.
Note that
  $\ket{\Psi_{\rm H}}$ and $\ket{\Psi_{\rm L}}$ are eigenvectors at
  each $\bk$.
Thus, we obtain
 $f_0(\bk) = (E_{\rm H}(\bk) + E_{\rm L}(\bk))/2$,
 $f_3(\bk) = (E_{\rm H}(\bk) - E_{\rm L}(\bk))/2$,
 and $f_2(\bk) = \bra{\Psi_{\rm H}} H_{\ho-\lo}\ket{\Psi_{\rm L}}$
  $ =   \bra{\Psi_{\rm L}} H_{\ho-\lo}\ket{\Psi_{\rm H}}$.
The quantities $f_0(\bk)$, $f_2(\bk)$, and $f_3(\bk)$
 can be taken as real.\cite{r26}
Note that
$f_0(\bk) = f_0(-\bk)$, and
$f_3(\bk) = f_3(-\bk)$ owing to  the time reversal symmetry
and
 $f_2(\bk) = -f_2(-\bk)$
 owing to the different parity of HOMO and LUMO.
The nodal line is obtained from   $f_3(\bk) = 0$
 and   $f_2(\bk)=0$.
In the next section,
 we examine these surfaces of
 $f_2(\bk)=0$ and $ f_3(\bk)=0$
 in the three-dimensional momentum space.

\section{Results}

\subsection{Band structure with  Dirac cone}
\begin{figure}
  \centering
\includegraphics[width=6cm]{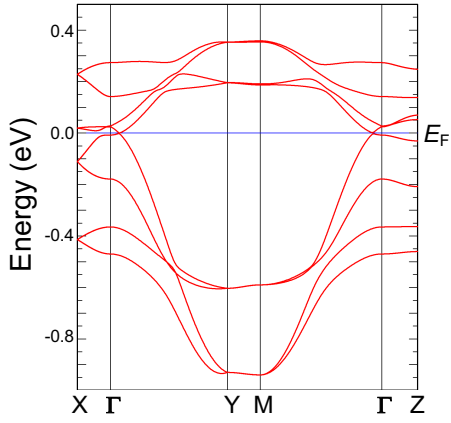}
  \caption{
 (Color online)
Energy band structure of [Pd(dddt)$_2$] at 5.9 GPa
}
\label{fig:fig2}
\end{figure}

In Fig. \ref{fig:fig2}, the energy band structure
is shown, where the origin of the energy is taken at
 the Fermi energy $E_{\rm F}$.
Despite  the rather large differences in transfer energies, the essential shape of the band structure, including the nodal line semimetal state, is very similar to that of the previous model.
The band crossing occurs on the line between TRIMs $\Gamma$ and Y.
The corresponding energy is slightly lower than the Fermi energy ($E_{\rm F}$)
 leading to an electron pocket.

\begin{figure}
  \centering
\includegraphics[width=8cm]{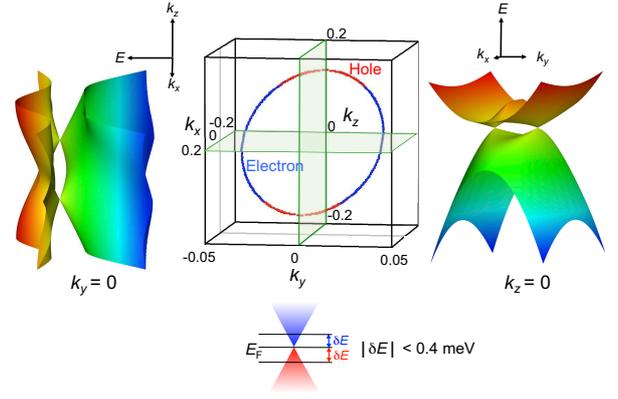}
  \caption{
 (Color online)
Nodal line and a pair of Dirac cones at $k_z$ = 0 and $k_y$ = 0 in [Pd(dddt)$_2$] at 5.9 GPa. The hole-like characteristic  is indicated in red and the electron-like characterisic in blue.
}
\label{fig:fig3}
\end{figure}

Here, we mention the Dirac point obtained from $E_g(\bkD) = 0$, i.e.,
$E_4(\bkD) = E_5(\bkD)$, which provides a line in the three-dimensional
 momentum space.
Figure \ref{fig:fig3} shows Dirac points forming a closed line
 (loop), which is symmetric with respect to $k_y=0$.
Compared with   the previous case in Ref. \citen{r11},
the loop is almost coplanar and located within the first Brillouin zone.
The  variation of the energy at the Dirac point along the line
 is very small, as shown in the next paragraph.

 Results of the calculation are summarized as follows.
The energy band of the 8 $\times$ 8 model in Fig. \ref{fig:fig2} reproduces the overall behavior of
the previous result that is obtained on the basis of the DFT.\cite{r10}
This is partially because the nature of the main transfer energies including $b1_{\rm H}$ and $b2_{\rm L}$ does not change.
Dirac points form a nodal loop within the first Brillouin zone (Fig. 3),
which well reproduces the nature of the DFT band structure.
The axis of the cone changes along the line, and the axis at $k_z=0$ is almost perpendicular to that at  $k_y=0$. The energy at the Dirac points varies along the nodal line but the deviation from the Fermi energy is very small,
 $\sim \pm 0.4$ meV, which is smaller (electron pocket) around $k_z=0$  and larger (hole pocket) around $k_y = 0$. Thus, the electronic state is expected to have a two-dimensional characteristic of the Dirac cone, as shown later in DOS.

\begin{figure}
  \centering
\includegraphics[width=8cm]{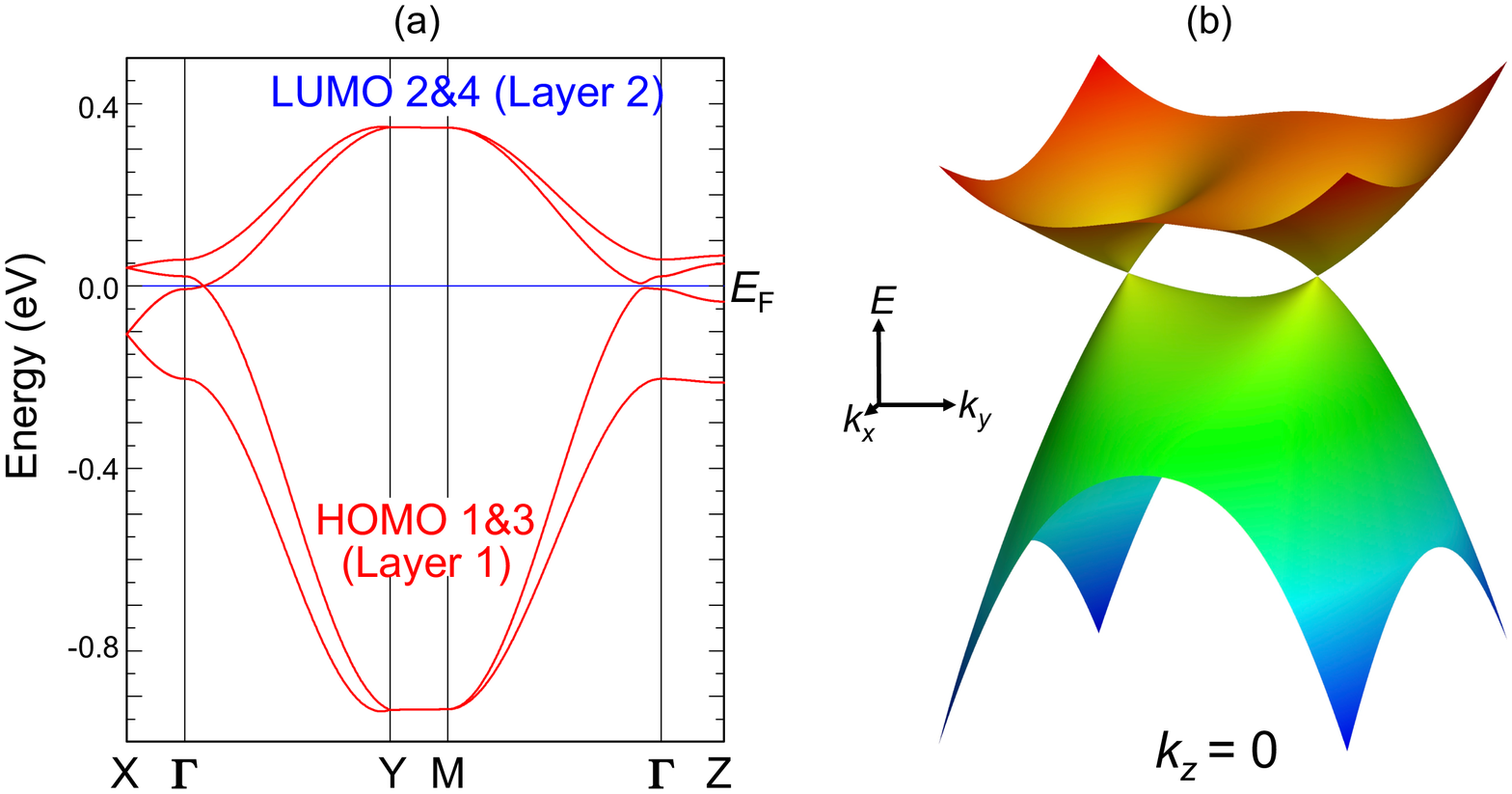}   \\
\includegraphics[width=8cm]{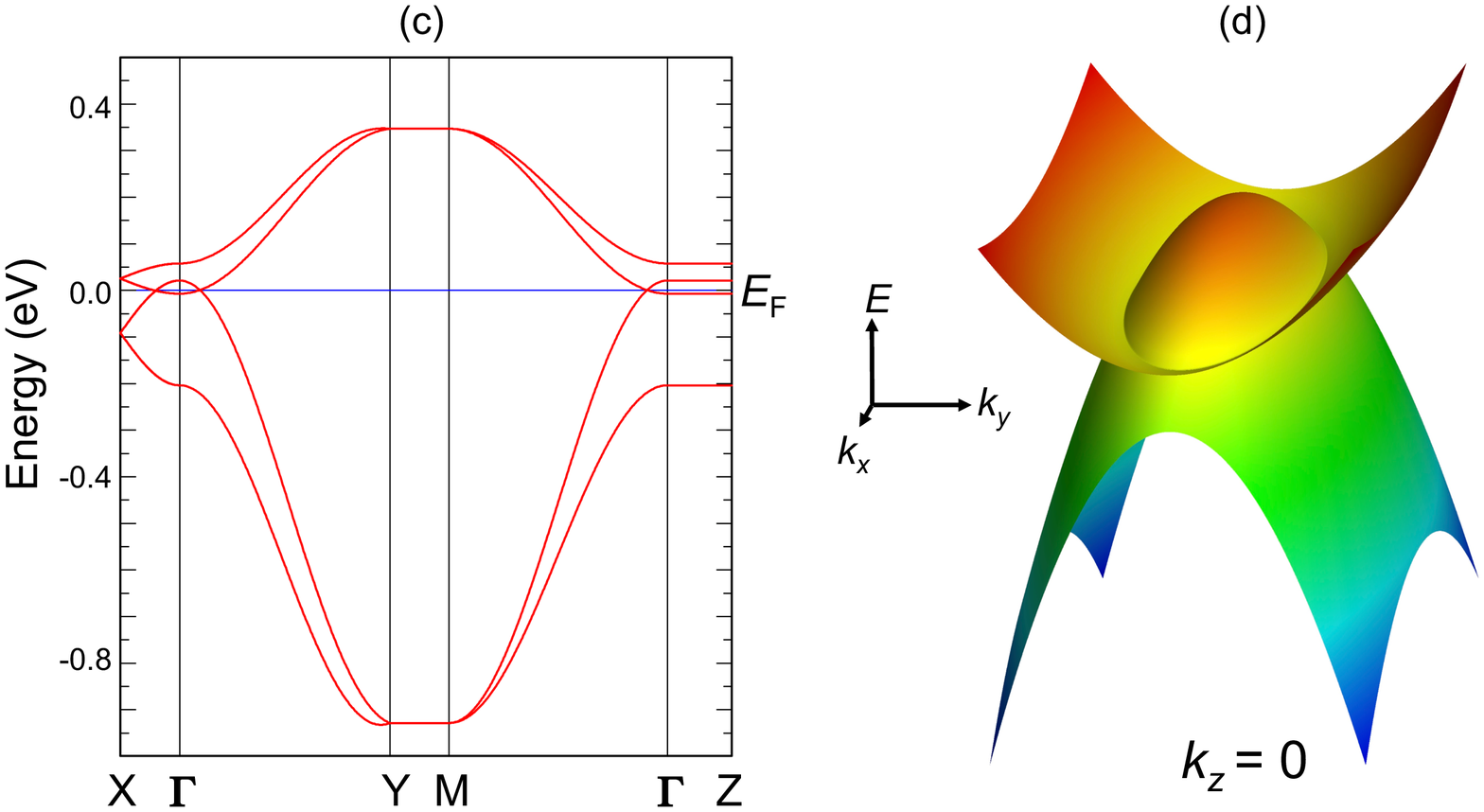}   \\
  \caption{
 (Color online)
Band dispersion (a) and Dirac cones at $k_z = 0$ (b) for the 4$\times$4 model. Band dispersion (c) and band energy dispersion surface at $k_z = 0$ for the 4$\times$4 model without HOMO--LUMO couplings ($h_{\rm H1,L2}=h_{\rm H1,L4}=h_{\rm H3,L2}=h_{\rm H3,L4}=0$) at $k_z = 0$ (d)
}
\label{fig:fig4}
\end{figure}

\begin{figure}
  \centering
\includegraphics[width=8cm]{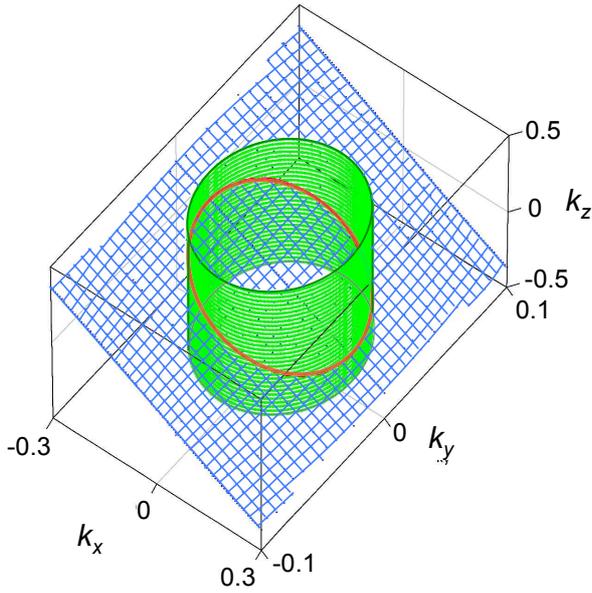}
  \caption{
 (Color online)
Intersection of HOMO and LUMO bands [$f_3(\bk)=0$: green cylinder], surface on which HOMO--LUMO couplings are zero [$f_2(\bk)=0$: blue plane], and nodal line (red line) in the 4$\times$4 model
}
\label{fig:fig5}
\end{figure}

 To consider a role of the direct interlayer HOMO--LUMO couplings in the Dirac cone formation, we examine  a reduced 4 $\times$ 4 Hamiltonian obtained using Eq. (\ref{eq:eq1})
by discarding the states
 $\ket{\HH2}, \ket{\HH4}, \ket{\LL1}$, and $\ket{\LL3}$, i.e.,
\begin{eqnarray}
H_{\rm red}
    & =&
   \sum_{\bk} \sum_{\alpha',\beta'} h_{\alpha', \beta'}(\bk)
 \ket{\bk, \alpha'}\bra{\bk, \beta'} \nonumber \\
  & =&
    \sum_{\bm{k}} \ket{\Phi_4(\bm{k})} \hat{H}_{ 4 \times 4}(\bm{k}) \bra{\Phi_4(\bm{k})}\; ,
 \label{eq:eq6}
 \end{eqnarray}
 with  $\alpha',\beta'$ = H1, H3, L2, and L4, and
($\bra{\Phi_4(\bm{k})} =  \bra{\HH1},\bra{\HH3},\bra{\LL2},\bra{\LL4}$).
This reduced model of the 4 $\times$ 4 Hamiltonian well reproduces the four energy bands shown in  Fig. \ref{fig:fig2} around the Fermi level
[Figs. \ref{fig:fig4}(a) and \ref{fig:fig4}(b)].
This means that the Dirac points originate from the HOMO bands in layer 1 and the LUMO bands in layer 2.
  Figures \ref{fig:fig4}(c) and \ref{fig:fig4}(d)
show that the Dirac cone vanishes  when the HOMO--LUMO couplings become zero ($h_{\rm H1,L2}=h_{\rm H1,L4}=h_{\rm H3,L2}=h_{\rm H3,L4}=0$).
Figures \ref{fig:fig4}(b) and \ref{fig:fig4}(d) show
 that the Dirac points (nodal line) emerge
 when the HOMO and LUMO bands meet  without the  HOMO--LUMO couplings on the
special points (the Dirac points). That is, the nodal line is an intersection of the surface of $f_3(\bk)=0$ and that of $f_2(\bk)=0$, each of which forms a cylinder and a plane in the three-dimensional momentum space, respectively (Fig.  \ref{fig:fig5}).
Note that Fig. \ref{fig:fig5} shows a qualitative behavior, since  the surfaces, i.e.,  $f_2(\bk) = 0$ and $f_3(\bk) = 0$, are evaluated by the perturbational method.

\begin{figure}
  \centering
\includegraphics[width=8cm]{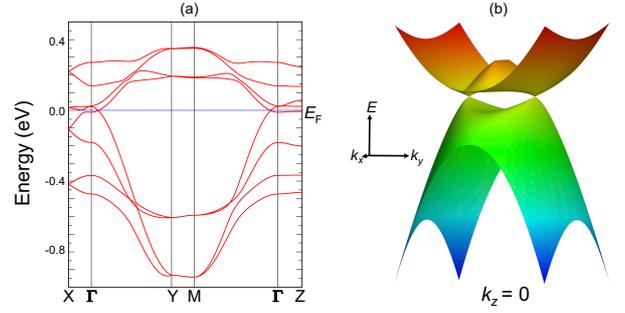}
  \caption{
 (Color online)
Band dispersion (a) and Dirac cones at $k_z$=0 (b) for the 8$\times$8 model in the case of $h_{\rm H1,L2}=h_{\rm H1,L4}=h_{\rm H3,L2}=h_{\rm H3,L4}=0$
}
\label{fig:fig6}
\end{figure}

Furthermore, the global band structure shown in Fig. \ref{fig:fig6}, in which the direct HOMO--LUMO couplings corresponding to the  4 $\times$ 4 model are discarded,  still resembles that shown in  Fig.~\ref{fig:fig2}.  This finding also suggests an additional mechanism of forming the Dirac point via a HOMO--LUMO interaction through a process of 2nd-order perturbation, as demonstrated in Ref. \citen{r10}. We return to this point later in the discussion of DOS.

\subsection{DOS}

We examine  DOS  $D(\omega)$
 per unit cell, which is defined as
\begin{equation}
D(\omega) = \frac{1}{N}
  \sum_{\bk} \sum_{\gamma }
 \delta (\omega - E_{\gamma}(\bk)) \; .
\label{eq:eq7}
\end{equation}
 A chemical potential $\mu$ corresponding  to
  a half-filled band
 is obtained from
 $ 4 = \int_{-\infty}^\mu d \omega \; D(\omega)$.

\begin{figure}
  \centering
\includegraphics[width=6cm]{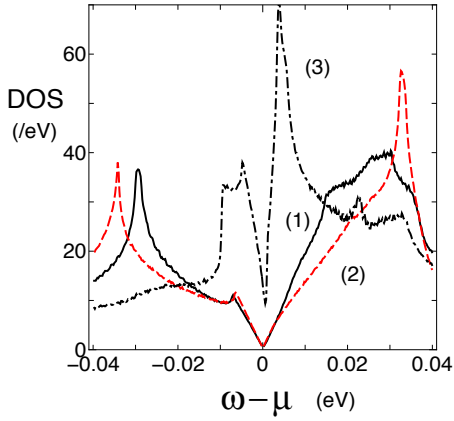}
\caption{(Color online)
DOS $D(\omega)$ per unit cell
for the 8 $\times$ 8 Hamiltonian (solid line) (1),
  and 4 $\times$ 4  Hamiltonian (dashed line) (2).
 The chemical potential $\mu$ is given by
 $\mu$  = 0.5053 for the 8 $\times$ 8  model and
 0.5095 for the 4 $\times$ 4 model.
Around $\omega - \mu$, the behavior $D(\omega) = K  |\omega - \mu|$
 with a constant $ K (> 0) $ is seen owing to a Dirac cone, where
 a small deviation close to $\omega - \mu$ comes from
 the energy variation on the nodal line.
The dot-dashed line (3) denotes
DOS without the interlayer H-L interaction, where $\mu$ = 0.5079.
}
\label{fig7}
\end{figure}

Figure \ref{fig7} shows DOS for both the  8 $\times$ 8 Hamiltonian [line (1)]
 and 4 $\times$ 4 Hamiltonian [line (2)], where the latter consists of  four bases, namely,  $\ket{\HH1},  \ket{\HH3}, \ket{\LL2}$, and  $\ket{\LL4}$.
 A good coincidence of DOS
  between the 8 $\times$ 8 and 4 $\times$ 4 Hamiltonians is found  for
 $-0.01 < \omega - \mu  < 0.005$.
  This suggests the validity of the 4 $\times$ 4  Hamiltonian
  to describe   the physical quantities at low temperatures.
The  difference in the global DOS between lines (1) and (2) is apparent
 for large $\omega - \mu( > 0)$, but
 the DOS for line (1) and that for line (2) are similar qualitatively,
 suggesting that the direct HOMO--LUMO interaction is
  crucial for the present Dirac electron system.
This can be understood by comparings
 the DOSs for lines (1) and (2)
  with  that for line (3), which is obtained
 by discarding the direct HOMO--LUMO interaction in
 the 8 $\times$ 8  Hamiltonian [Eq. (\ref{eq:eq1}].
Line (3) corresponds to the energy band shown in Fig. \ref{fig:fig6}, and is similar to that in a previous paper. \cite{r12}
 There is  a qualitative difference in DOS
   between  lines (1) and  (3) in the sense that
 the linear dependence of DOS around $\omega = \mu$
 in lines (1) and (2)
 suggests  an almost   zero-gap state (ZGS) and is robust, i.e., displaying   a wide energy region of
linear dependence owing  to the direct HOMO--LUMO interaction, which
 is in contrast to  that in line (3)  without
 the direct HOMO--LUMO interaction.
A detailed analysis close to $\omega = \mu$   shows that
 the deviation of $D(\omega)$ from that with
the linear dependence occurs in the narrow region of
$|\omega - \mu| < 0.001$, suggesting
 that the  variation of the energy on the nodal line is less than  0.001.

Here, we note that such a linear dependence of DOS has been
 found in the two-dimensional organic conductor
 $\alpha$-ET$_2$I$_3$,
 which is the first material of Dirac electrons
 in a molecular conductor.
\cite{Kobayashi2004,Katayama2006_Dirac,Katayama_EPJ}
The similarity of DOS for {Pd(dddt)$_2$] and that for \ET
 suggests that such a three-dimensional nodal line semimetal shares
  a common feature with the two-dimensional Dirac cone.

Moreover, we note that the linear dependence is given by
 $D(\omega) = K |\omega - \mu|$ with a coefficient
 $K \propto  v^{-2}$, where
  $v$ denotes an average  velocity of the  Dirac cone.
Since $K$ of  lines (1) and (2) is smaller than that of line (3),
 the former velocity is larger than the latter one, suggesting
 that  in Eq. (\ref{eq:eq4}), $f_2(\bk)$ of the former is larger than that
of the latter. This is reasonable since the direct HOMO--LUMO interaction
 is present in lines (1) and (2)
 but is absent  in line (3).

\subsection{Parity at TRIM}
To analyze the  Dirac point,
  we calculate the parity at the TRIM
 given by $\bm{G}$/2 with $\bm{G}$ being the reciprocal lattice vector,
 where
  $\bm{G}/2 =(0,0,0)$, $(1/2, 0 , 0)$,  $ (0, 1/2 ,0)$,
 and  $(1/2,1/2,0)$  correspond to the $\Gamma$, X, Y, and M points,
and
 $\bm{G}/2 =(0, 0, 1/2)$, $(1/2, 0, 1/2)$,  $ (0,1/2, 1/2)$, and
$(1/2, 1/2, 1/2)$
 correspond to the
  Z, D, C, and E points, respectively.

 The inversion with respect to a Pd atom of
   molecule 1 in the crystal structure gives
the matrix for  the  translation of the base
  (H1, H2, $\cdots$, L4),  $\hat{P}(\bk)$,
  expressed as\cite{r11}
\begin{eqnarray}
 \hat{P}(\bk) & =  &
\begin{pmatrix}
 \hat{P}_1(\bk) & 0   \\
  0 & -  \hat{P}_1(\bk)
 \\
 \end{pmatrix} \; ,
\label{eq:eq8}
\end{eqnarray}
where $\hat{P}_1(\bk)$ denotes a 4 $\times 4$ matrix,
\begin{eqnarray}
 \hat{P}_1(\bk) & =  &
\begin{pmatrix}
 - 1 & 0 & 0 & 0  \\
  0 & -XYZ & 0 & 0 \\
  0 & 0 & -XY & 0          \\
  0 & 0 & 0 & -\bar{Z}  \\
 \end{pmatrix} .
 \nonumber \\
\label{eq:eq9}
\end{eqnarray}
The relation
$(\hat{P}(\bm{k}))_{Hj,Hj}
= - (\hat{P}(\bm{k}))_{Lj,Lj}$
for $j$=1, 2, 3, and 4
 comes from the fact that the HOMO has an ungerade symmetry and the LUMO has a gerade symmetry.
The eigenvalue and eigenfunction of $\hat{P}(\bk)$
  are obtained from  ($\alpha$ = H1, H2, $\cdots$, L4)
\begin{equation}
\hat{P}(\bm{k}) u_\alpha(\bk)
 = p_\alpha(\bk) u_\alpha(\bk) \; ,
\label{eq:eq10}
\end{equation}
where
 $p_\alpha(\bk) =  (\hat{P}(\bm{k}))_{\alpha,\alpha }$,
   $u_{\rm H1}(\bk) =u_1 = (1,0,0,0,0,0,0,0)^t$,
  $u_{\rm H2} = u_2 =(0,1,0,0,0,0,0,0)^t$,
 $\cdots$, and
  $u_{\rm L4} = u_8= (0,0,0,0,0,0,0,1)^t$.
At the TRIM,  one obtains  $p_\alpha(\bm{G}/2) = p_l = + (-)$,
  which gives an  even (odd) parity.
The parity $p_l = p_l(\hg)$ is listed in Table \ref{table_1}.
From Eq. (\ref{eq:eq8}), it is obvious that  $\sum_{l,\bm{G}} p_l(\hg) = 0$,
 i.e.,
 the number of the even parities is the same as that of the odd parities.

Since $[\hat{P}(\bm{G}/2), \hat{H}(\bm{G}/2)] = 0$,
   $\Psi_j(\bm{G}/2)$ in Eq. (\ref{eq:eq2})
is also an eigenfunction of
 $\hat{P}(\bm{G}/2)$.
  The corresponding equation at the TRIM is given by
\begin{eqnarray}
 \label{eq:eq11}
 \hat{P}(\bm{G}/2) \Psi_j(\bm{G}/2) =
      E_{P}(j,\hg) \Psi_j(\bm{G}/2)
 \; ,
\end{eqnarray}
 with  $E_P(j,\hg) = + (-)$,
 which denotes an even (odd) parity.
In terms of $u_l(\hg)$, $\Psi_j(\hg)$ is expressed as
\begin{eqnarray}
 \label{eq:eq12}
 \Psi_j(\hg) = \sum_{l} d_{j,l}(\hg)   u_{l}(\hg) \; .
\end{eqnarray}
 Since $\hat{P}(\hg) u_l(\hg) = p_l(\hg) u_l(\hg)$ with
 $ p_l(\hg) = \pm$, it turns out that
 $d_{j,l}(\hg)$ = 0 for $p_l(\hg) \not= E_P(j,\hg)$.
 Thus, the wave function for $E_j(\hg)$ is given by
    a linear combination of the base $u_l(\hg)$,
   which has the same parity as $E_P(j,\hg)$.
From Eqs. (\ref{eq:eq11}) and (\ref{eq:eq12}),
 $E_P(j,\hg)$ is estimated   as
\begin{eqnarray}
 E_P(j,\hg) &=&
    \Psi_j(\bm{G}/2)^\dagger \hat{P}(\bm{G}/2 ) \Psi_j(\bm{G}/2)
 \nonumber \\
 &=& \sum_{l = 1}^{8} p_{l}(\bm{G}/2) |d_{j,l}|^2 \; ,
 \label{eq:eq13}
\end{eqnarray}
which is utilized for the present numerical calculation.
$E_P(j,\hg)$ is shown
as a function of $E_j(\hg)$
for the respective TRIM.
The sign of some elements
in Table \ref{table2}
 is different from that of
a  previous calculation.~\cite{r11}} However, as shown later, the resultant conditions for the Dirac points  are
 unchanged,  since both share a common feature of  a nodal line.
 Note that $\sum_j P_{E_j}(\hg)$  = 0 for the respective
 TRIM.
 For the  wave function $\Psi_j(\bm{G}/2)$ in Eq. (\ref{eq:eq11})
 at the  $\Gamma$ and M points,
 the even (odd) parity corresponds to  LUMO (HOMO).


To examine the nodal loop between $E_4(\bk)$ and $E_5(\bk)$,
  we calculate
$P_\delta$ [$\delta$ =1
 for  $k_z=0$, $\delta$ =2 for $k_y=0$,
and $\delta$ =3  for $k_z =0.5$], which is
  defined as\cite{r11}
\begin{subequations}
 \label{eq:2D_parity}
\begin{eqnarray}
P_1 & = & \prod_{j=5}^{8}
 E_{P}(j,\Gamma)E_P(j,{\rm X}) E_P(j,{\rm Y})E_P(j,{\rm M}) \; ,
\label{eq:eq14a}
\nonumber  \\ \\
 P_2
   &=& \prod_{j=5}^{8}
 E_{P}(j,{\rm Z})E_{P}(j,\Gamma) E_{P}(j,{\rm X})E_{P}(j,{\rm D}) \; ,
 \label{eq:eq14b} \nonumber  \\  \\
 P_3
   &=& \prod_{j=5}^{8}
 E_{P}(j,{\rm Z})E_{P}(j,{\rm D}) E_{P}(j, {\rm E})E_{P}(j,{\rm C}) \; .
 \label{eq:eq14c} \nonumber  \\
\end{eqnarray}
\end{subequations}
  Since we examine the Dirac point of the level crossing
 between $E_4$ and $E_5$ bands owing to a half-filled band,
 we take $j$ = 5, 6, 7, and 8.
Each $P_{\delta}$ denotes a quantity  assigned on a plane including
the four corresponding    TRIMs.
The condition for the Dirac point  between $E_4$ and $E_5$
 is given by $P_j= \pm 1$.~\cite{Herring1937}
 When  $P_j = -1$, the number of pairs of Dirac points
  between $E_4$ and $E_5$ is odd,~\cite{r24,r25}
 implying that the topological number is 1.
The condition for a nodal line is given by~\cite{r11}
\begin{equation}
 P= P_1 P_3 = -1 \; .
\label{eq:eq15}
\end{equation}
Note that Eqs.
(\ref{eq:eq14a})--(\ref{eq:eq14c})
 describe  the condition of the Dirac point  on the planes of $k_z = 0$ (TRIM with the $\Gamma$, X, Y, and M points),
 $k_y = 0$   (TRIM with the Z, $\Gamma$, X, and D points), and
 $k_z = 0.5$ (TRIM with the Z, D, C, and  E points), respectively.
The nodal closed loop is found for $P_1 = -1$,  $P_2 = -1$, and
  $P_3 = 1$.
Note that the parity for the nodal line semimetal
  of [Pt(dmdt)$_2$]
 with an open line \cite{Zhou2019}
 is given by $P$ = 1 \cite{Kato2020_JPSJ}
 instead of  Eq. (\ref{eq:eq15}).

\begin{table}
\caption{
 Parity  $p_l(\hg)$, where
$\hat{P}({\bf G}/2) u_l(\hg) = p_l(\hg) u_l(\hg)$.
 }
\begin{center}
\begin{tabular}{cccccccccc}
\hline\noalign{\smallskip}
$$ &  $u_1$ & $u_2$ & $u_3$ & $u_4$
                               &  $u_5$ & $u_6$ & $u_7$ & $u_8$  \\
\noalign{\smallskip}\hline\noalign{\smallskip}
$ \Gamma$   & $-$  & $-$  & $-$  & $-$     & $+$  & $+$  & $+$  & $+$   \\
$ {\rm X}$  & $-$  & $+$  & $+$  & $-$     & $+$  & $-$  & $-$  & $+$   \\
$ {\rm Y}$  & $-$  & $+$  & $+$  & $-$     & $+$  & $-$  & $-$  & $+$   \\
$ {\rm M}$  & $-$  & $-$  & $-$  & $-$     & $+$  & $+$  & $+$  & $+$   \\
$ {\rm Z}$  & $-$  & $+$  & $-$  & $+$     & $+$  & $-$  & $+$  & $-$   \\
$ {\rm D}$  & $-$  & $-$  & $+$  & $+$     & $+$  & $+$  & $-$  & $-$   \\
$ {\rm C}$  & $-$  & $-$  & $+$  & $+$     & $+$  & $+$  & $-$  & $-$   \\
$ {\rm E}$  & $-$  & $+$  & $-$  & $+$     & $+$  & $-$  & $+$  & $-$   \\
\noalign{\smallskip}\hline
\end{tabular}
\end{center}
\label{table_1}
\end{table}

\begin{table}
\caption{
 Parity  $ E_P(j,\bm{G}/2) (= \pm) $
 of  8 x 8 Hamiltonian at P = 5.9 GPa as the function of
 $E_j (j =1,\dots 8)$ and $\hg$ (= $\Gamma, \cdots$,  E).
 }
\begin{center}
\begin{tabular}{cccccccccc}
\hline\noalign{\smallskip}
$$ &  $E_1$ & $E_2$ & $E_3$ & $E_4$
                                  &  $E_5$ & $E_6$ & $E_7$ & $E_8$  \\
\noalign{\smallskip}\hline\noalign{\smallskip}
$ \Gamma$   & $\m$  & $\m$  & $\p$  & $\m$     & $\m$  & $\p$  & $\p$  & $\p$   \\
X   & $\m$  & $\p$  & $\m$  & $\p$     & $\p$  & $\m$  & $\m$  & $\p$   \\
Y  & $\m$  & $\p$  & $\m$  & $\p$     & $\p$  & $\m$  & $\p$  & $\m$   \\
M   & $\m$  & $\m$  & $\m$  & $\m$     & $\p$  & $\p$  & $\p$  & $\p$   \\
 Z   & $\m$  & $\m$  & $\p$  & $\p$     & $\p$  & $\p$  & $\m$  & $\m$   \\
D   & $\m$  & $\p$  & $\m$  & $\p$     & $\p$  & $\m$  & $\p$  & $\m$   \\
C   & $\m$  & $\p$  & $\m$  & $\p$     & $\p$  & $\m$  & $\p$  & $\m$   \\
E  & $\p$  & $\p$  & $\m$  & $\m$     & $\m$  & $\m$  & $\p$  & $\p$   \\
\noalign{\smallskip}\hline
\end{tabular}
\end{center}
\label{table2}
\end{table}

Here, we examine the reduced 4 $\times$ 4 Hamiltonian with the bases
 of H1, H3, L2, and L4 by discarding the elements corresponding to
  H2, H4, L1, and L3.
Instead of eight bands, we obtain the four bands $E_1(\bk)$, $E_2(\bk)$,
 $E_3(\bk)$, and $E_4(\bk)$.
In this case, the inversion matrix is given by
\begin{eqnarray}
 \hat{P}_{4 \times 4}(\bk) & =  &
\begin{pmatrix}
 - 1 & 0 & 0 & 0  \\
  0 & -XY & 0 & 0 \\
  0 & 0 & XYZ & 0          \\
  0 & 0 & 0 & \bar{Z}  \\
 \end{pmatrix} .
 \nonumber \\
\label{eq:eq16}
\end{eqnarray}
The parities are listed in Table \ref{table3}.
Noting that the filled band is given by $E_j$ with $j$=3 and 4,
we obtain also the closed nodal line
 since  $P(k_z=0) = -1$,  $P(k_y=0) = -1$, and
  $P(k_z=\pi) = 1$.

Here,
we comment on the parity of the 4 $\times$ 4 reduced Hamiltonian,
 which is different from the conventional one. \cite{r25}.
The relation  $\sum_{l,\bm{G}} p_l(\hg) = 0$ holds for the latter case
 but does not for $\hg$ = Z and E of the former,
 in which the Pd atom is not the inversion center
 because H2, H4, L1, and L3 are discarded.
Thus, we construct
 an effective  8 $\times$ 8 Hamiltonian
 by adding 4 bases, namely,  H2, H4, L1, and L3 with only  site energies,
 which are
 much higher (lower)
 than  $E_j$  $(j=1,\cdots,4)$ for
  H2 and H4 (L1 and L3).
In this case, the energies are obtained as
$E_a > E_b \gg E_1 >E_2 >E_3 >E_4  \gg E_c > E_d$,
where $E_1, \cdots, E_4$ are the same as those of the  4  $\times$ 4 Hamiltonian.
We obtain the additional parity as
  [$E_P(j,\Gamma)$,
   $E_P(j,{\rm X})$,
   $E_P(j,{\rm Y})$,
   $E_P(j,{\rm M})$,
   $E_P(j,{\rm Z})$,
   $E_P(j,{\rm D})$,
   $E_P(j,{\rm C})$,
   $E_P(j,{\rm E})$]
 =
 $(\p,\p,\p,\p,\p,\p,\p,\p)$ for $E_j = E_a$,
 $(\p,\m,\m,\p,\p,\m,\m,\p)$ for $E_j= E_b$,
 $(\m,\p,\p,\m,\p,\m,\p,\m)$ for $E_j = E_c$, and
 $(\m,\m,\m,\m,\p,\p,\p,\p)$ for $E_j= E_d$.
 We find  that  the resultant parity  satisfies
$\sum_{l,\bm{G}} p_l(\hg) = 0$, and the parity  relevant to
  $E_1, E_2, E_3$, and $E_4$ remains the same
  as that in Table~\ref{table3}.

\begin{table}
\caption{
 Parity  $ E_{P}(j,\bm{G}/2) (= \pm) $
 of  4 x 4 Hamiltonian with H1, H3, L2, and L4.
}
\begin{center}
\begin{tabular}{cccccc}
\hline\noalign{\smallskip}
$$ &  $E_1$ & $E_2$
                                  &  $E_3$ & $E_4$   \\
\noalign{\smallskip}\hline\noalign{\smallskip}
$\Gamma$   &  $\m$  & $\p$     & $\m$  & $\p$     \\
${\rm X}$  &  $\p$  & $\m$     & $\p$  & $\m$     \\
${\rm Y}$  &  $\p$  & $\m$     & $\p$  & $\m$     \\
${\rm M}$  &  $\m$  & $\m$     & $\p$  & $\p$     \\
${\rm Z}$  &  $\p$  & $\p$     & $\p$  & $\p$     \\
${\rm D}$  &  $\p$  & $\m$     & $\p$  & $\m$     \\
${\rm C}$  &  $\p$  & $\m$     & $\p$  & $\m$     \\
${\rm E}$  &  $\p$  & $\p$     & $\p$  & $\p$   \\
\noalign{\smallskip}\hline
\end{tabular}
\end{center}
\label{table3}
\end{table}

\section{Conclusions}
We have examined a nodal line semimetal in a single-component molecular conductor
[Pd(dddt)$_2$] under high pressure, which consists of two crystallographically independent molecular layers.
On the basis of the synchrotron X-ray diffraction measurements at 5.9 GPa, we derive the TB model with both 8 $\times$ 8 and $4 \times 4$ matrix Hamiltonians.
We have shown a mechanism of the nodal line formation, which is obtained as an intersection between a crossing plane of the HOMO and LUMO bands and a plane of vanishing HOMO--LUMO interactions.
Compared with our previous paper, our new finding is the  crucial role of the direct HOMO--LUMO interaction between layers 1 and 2,
 which results in the robust Dirac cone
  within an energy height of $\sim$  0.01 eV.
This finding was verified by calculating DOS, which also provides a width of energy band along the nodal line of less than 0.4 meV.  Finally, in terms of topology, we examined the parity for the nodal line in both 8 $\times$ 8 and $4 \times 4$ matrix Hamiltonians. The former Hamiltonian is essentially the same as the previous one, wheras the latter is different from the conventional $4 \times 4$ matrix Hamiltonian but can be interpreted in a consistent manner.

\acknowledgements
We acknowledge Diamond Light Source for time on Beamline I19-2 under Proposal MT20934-1 and we thank Dr. Dave Allan and Dr. Lucy Saunders for their assistance.
This work was supported by JSPS KAKENHI (Grant no.  JP16H06346).

\appendix

\section{Crystal data of [Pd(dddt)$_2$] at 5.9 GPa}

\noindent
Space group: $P2_{1}/n$ \\
Lattice constants: $a_{\rm o} = 16.48,\ b_{\rm o} = 4.3102,\ c_{\rm o} = 17.480$  \AA, $\beta_{\rm o}=111.84^\circ $

From  these data, fractional atomic coordinates
are shown in Table \ref{table4}. \\

In the calculation, we define the following new cell where the $a$-axis is parallel to layers 1 and  2: \\
$ \bm{a} = -(\bm{a}_{\rm o}+\bm{c}_{\rm o}), \bm{b} = -\bm{b}_{\rm o}, \bm{c} = \bm{c}_{\rm o} $ \\

\begin{table}
\caption{
 Fractional atomic coordinates.\
 }
\begin{center}
\begin{tabular}{crrr}
\hline\noalign{\smallskip}

  \multicolumn{1}{c}{Atom} &
 \multicolumn{1}{c}{$x$} &
 \multicolumn{1}{c}{$y$} &
 \multicolumn{1}{c}{$z$} \\

\noalign{\smallskip}\hline\noalign{\smallskip}
Pd1  & 0.00000  &  0.00000   & 0.00000    \\
S1  &  0.10040  & $-$0.1850  & $-$0.04520 \\
S2  &  0.06100  & $-$0.2850  & 0.11400    \\
S3  &  0.24700  & $-$0.5960  & 0.00620    \\
S4  &  0.20400  & $-$0.7030  & 0.18400    \\
C1  &  0.15600  & $-$0.42706 & 0.02800    \\
C2  &  0.14800  & $-$0.4640  & 0.10100    \\
C3  &  0.30000  & $-$0.8460  & 0.09100    \\
C4  &  0.31100  & $-$0.6930  & 0.17600    \\
H3A &  0.26727  & $-$1.0371  & 0.084043   \\
H3B &  0.35779  & $-$0.89871 & 0.091206   \\
H4A &  0.33099  & $-$0.48076 & 0.17840    \\
H4B &  0.35293  & $-$0.80806 & 0.22127	  \\
Pd2 &  0.50000  &  0.5000 & 0.00000 \\
S5  & 0.60930   &  0.8400 & 0.05810 \\
S6  & 0.45870   &  0.5950 & 0.10490 \\
S7  & 0.67580   &  1.2230 & 0.20240 \\
S8  & 0.50200   &  0.9570 & 0.24860 \\
C5  & 0.60300   &  0.9530 & 0.14700 \\
C6  & 0.52900   &  0.8510 & 0.16600 \\
C7  & 0.67900   &  1.1850 & 0.30800 \\
C8  & 0.58400   &  1.2150 & 0.30300 \\
H7A & 0.70275   &  0.9846 & 0.33026 \\
H7B & 0.71493   &  1.3461 & 0.34237 \\
H8A & 0.56407   &  1.4195 & 0.28101 \\
H8B & 0.58671   &  1.2140 & 0.35933 \\
\noalign{\smallskip}\hline
\end{tabular}
\end{center}
\label{table4}
\end{table}

\section{Matrix elements of Hamiltonian}
The matrix elements for HOMO--HOMO (H--H) are given by
\begin{eqnarray}
h_{\rm H1,H1}&=& b1_{\rm H}(Y+\bar{Y}) \nonumber
\; , \\
h_{\rm H1,H2}&=& a1_{\rm H}(XZ+Y)+a2_{\rm H}(1+XYZ)  \nonumber
\; , \\
h_{\rm H1,H3}&=& p_{\rm H}(1+X+Y+XY)  \nonumber
\; , \\
h_{\rm H1,H4}&=& c1_{\rm H}(1+\bZ)+c2_{\rm H}(Y+\bY \bZ)
\; ,  \nonumber \\
h_{\rm H2,H2}&=& b2_{\rm H}(Y+\bY)
\; , \nonumber \\
h_{\rm H2,H3}&=& c1_{\rm H}(1+\bZ)+c2_{\rm H}(\bY+ Y \bZ)
\; , \nonumber \\
h_{\rm H2,H4}&=& q_{\rm H}(\bX\bZ+\bY\bZ+\bX\bY\bZ+\bZ)
 \nonumber \; , \nonumber \\
h_{\rm H3,H3}&=& b1_{\rm H}(Y+\bY)
\; , \nonumber \\
h_{\rm H3,H4}&=& a1_{\rm H}(1+\bX\bY\bZ) + a2_{\rm H}(\bX\bZ+\bY)
\; , \nonumber \\
h_{\rm H4,H4}&=& b2_{\rm H}(Y+\bY)
\; , \nonumber 
\end{eqnarray}
The matrix elements for HOMO--LUMO (H-L) are given by
\begin{eqnarray}
h_{\rm H1,L1}&=&  b1_{\rm HL}(\bY - Y)
\; , \nonumber\\
h_{\rm H1,L2}&=&  a1_{\rm HL}(Y-XZ)+a2_{\rm HL}(1-XYZ)
\; , \nonumber \\
h_{\rm H1,L3}&=&  p1_{\rm HL}(1-XY)+ p2_{\rm HL}(Y-X)
\; , \nonumber\\
h_{\rm H1,L4}&=&  c1_{\rm HL}(1-\bZ)+c2_{\rm HL}(Y-\bY\bZ)
\; , \nonumber\\
h_{\rm H2,L1}&=&  a1_{\rm HL}(\bX\bZ-\bY)+a2_{\rm HL}(1-\bX\bY\bZ)
\; ,\nonumber \\
h_{\rm H2,L2}&=&  b2_{\rm HL}(\bY-Y)
\; , \nonumber\\
h_{\rm H2,L3}&=&  c1_{\rm HL}(\bZ-1)+c2_{\rm HL}(\bY-Y\bZ)
\; , \nonumber\\
h_{\rm H2,L4}&=&  q1_{\rm HL}(\bX\bY\bZ-\bZ)+q2_{\rm HL}(\bX\bZ-\bY\bZ)
\; , \nonumber \\
h_{\rm H3,L1}&=&  p1_{\rm HL}(\bX-\bY)+p2_{\rm HL}(\bX\bY-1)
\; , \nonumber\\
h_{\rm H3,L2}&=&  c1_{\rm HL}(1-\bZ)+c2_{\rm HL}(Y-\bY Z)
\; , \nonumber\\
h_{\rm H3,L3}&=&  b1_{\rm HL}(-\bY+Y)
\; , \nonumber\\
h_{\rm H3,L4}&=&  a1_{\rm HL}(1-\bX\bY\bZ)+a2_{\rm HL}(\bY-\bX\bZ)
\; , \nonumber \\
h_{\rm H4,L1}&=&  c1_{\rm HL}(1-Z)+c2_{\rm HL}(YZ-\bY)
\; , \nonumber\\
h_{\rm H4,L2}&=&  q1_{\rm HL}(XZ-YZ)+ q2_{\rm HL}(XYZ-Z)
\; ,\nonumber \\
h_{\rm H4,L3}&=&  a1_{\rm HL}(1-XYZ)+a2_{\rm HL}(XZ-Y)
\; , \nonumber \\
h_{\rm H4,L4}&=&  b2_{\rm HL}(\bY-Y)
\; ,\nonumber 
\end{eqnarray}
The matrix elements for LUMO-LUMO (L-L) are given by
\begin{eqnarray}
h_{\rm L1,L1}&=&  \Delta E + b1_{\rm L}(Y+\bY)
\; , \nonumber\\
h_{\rm L1,L2}&=&  a1_{\rm L}(XZ+Y)+a2_{\rm L}(1+XYZ)
\; , \nonumber\\
h_{\rm L1,L3}&=&  p_{\rm L}(1+X+Y+XY)
\; , \nonumber\\
h_{\rm L1,L4}&=&  c1_{\rm L}(1+\bZ)+c2_{\rm L}(Y+\bY\bZ)
\; , \nonumber\\
h_{\rm L2,L2}&=&  \Delta E + b2_{\rm L}(Y+\bY)
\; ,\nonumber \\
h_{\rm L2,L3}&=& -c1_{\rm L}(1+\bZ)-c2_{\rm L}(\bY+Y\bZ)
\; , \nonumber\\
h_{\rm L2,L4}&=&  q_{\rm L}(\bX\bZ+\bY\bZ + \bX\bY\bZ + \bZ)
\; , \nonumber\\
h_{\rm L3,L3}&=&  \Delta E +b1_{\rm L}(Y+\bY)
\; ,\nonumber \\
h_{\rm L3,L4}&=& -a1_{\rm L}(1+\bX\bY\bZ) -a2_{\rm L}(\bX\bZ+ \bY)
\; , \nonumber \\
h_{\rm L4,L4}&=&  \Delta E + b2_{\rm L}(Y + \bY)
\; . \nonumber
\end{eqnarray}


\end{document}